# Orthogonality Catastrophes in Quantum Electrodynamics

R. Merlin

*Harrison M. Randall Laboratory of Physics, University of Michigan, Ann Arbor, MI 48109-1040*

The insertion of a small polarizable particle in an arbitrarily large optical cavity significantly alters the quantum-mechanical state of the electromagnetic field in that the photon ground state of the empty cavity and that of the cavity with the particle become mutually orthogonal and, thus, cannot be connected adiabatically in the infinite limit. The photon problem can be mapped exactly onto that of a many-body system of fermions, which is known to exhibit an orthogonality catastrophe when a finite-range local potential is introduced. We predict that the motion of polarizable objects inside a cavity, no matter how slow, as well as their addition and removal from the cavity, will generate a macroscopic, diverging number of low-energy photons. The significance of these results in regard to the quantum measurement problem and the dynamical Casimir effect are also discussed.

Quantum systems with an infinite number of degrees of freedom differ substantially from those with a finite number of variables in that they can be described alternatively by mutually orthogonal and, thus, inequivalent Hilbert spaces [1]. This well-known feature of quantum field theory is exemplified by the unitarily inequivalent representations resulting from the application of Bogoliubov-type transformations, which are central to many problems involving spontaneous symmetry breaking and, in particular, to the Higgs and BCS mechanisms for mass generation and superconductivity. Somehow less known outside condensed matter theory, is that a weak local potential can have a similar effect on a many-body system, as the overlap between the unperturbed and the ground state in the presence of the potential can vanish in the thermodynamic limit. This orthogonality catastrophe [2], broadly related to an infrared divergence, has been extensively studied for fermions (electrons) as it plays a crucial role in the understanding of the x-ray edge singularity in metals [3,4] and the Kondo problem [5]. Here we show that a closely related catastrophe can occur for photons in a cavity. Our approach distinguishes itself in many respects from the few, previously proposed boson (phonon) models exhibiting infrared divergences [6,7,8], all of which rely on chemical-bond displacements and depend quite sensitively on their long-wavelength behavior to produce the catastrophe.

Consider an arbitrarily-shaped cavity of volume $V$, partially filled with inclusions, which occupy a small volume $\upsilon << V$ and are made of one or more substances, all assumed to be isotropic, non-magnetic and lossless, so that the permeability is $\mu = 1$ everywhere, whereas the permittivity $\varepsilon$ depends both on frequency $\omega$ and position $\mathbf{r}$ and, for simplicity, is assumed to be real. Classically, the energy associated with a single mode of the cavity is [9]

$$\langle \mathcal{H}_\omega \rangle = \frac{1}{16\pi} \int_V \left\{ \frac{d[\omega\varepsilon(\mathbf{r},\omega)]}{d\omega} \mathbf{E}_\omega^2(\mathbf{r}) + \mathbf{B}_\omega^2(\mathbf{r}) \right\} d^3\mathbf{r} \tag{1}$$

where $\omega$ is the mode frequency, $\mathbf{E}_\omega$ and $\mathbf{B}_\omega$ are the electric and magnetic field and $\langle ... \rangle$ denotes time average (we recall that one must consider an infinitesimally narrow range of frequencies about $\omega$ to derive this expression [9]). The fields are assumed to have a time-dependence of the form $\exp(-i\omega t)$, which is hereafter omitted. Let $\Phi$ and $\mathbf{A}_\omega$ be, respectively the scalar and vector potentials; $\mathbf{B}_\omega = \nabla \times \mathbf{A}_\omega$. Choosing the generalized Coulomb gauge where $\Phi = 0$, we get $\mathbf{E}_\omega = -i\omega c^{-1}\mathbf{A}_\omega$ so that $\varepsilon(\mathbf{r},\omega)\omega^2 c^{-2}\mathbf{A}_\omega(\mathbf{r}) - \nabla \times \left[\nabla \times \mathbf{A}_\omega(\mathbf{r})\right] = 0$. After integration by parts, Eq. (1) becomes

$$\langle \mathcal{H}_\omega \rangle = \frac{\omega^2}{16\pi c^2} \int_V \left[ \frac{d(\omega\varepsilon)}{d\omega} + \varepsilon \right] \mathbf{A}_\omega^2(\mathbf{r}) d^3\mathbf{r} \quad . \tag{2}$$

The gauge is fixed by imposing the transversality condition $\nabla.[\varepsilon(\mathbf{r})\mathbf{A}_\omega] = 0$.

The first step in the quantization of the theory is the search for a classical Lagrangian that is consistent with both, the Hamiltonian, Eq. (1), and Maxwell's equation for $\mathbf{A}_\omega$ [10 - 19], a problem that is rather involved for a medium that is both dispersive and inhomogeneous [13 - 19]. Instead of pursuing a step-by-step path, we follow the heuristic, shortcut approach described in [15] and write $\mathbf{A}_\omega(\mathbf{r}) = C_\omega Q_\omega \mathbf{g}_\omega(\mathbf{r})$ where

$$C_\omega = \sqrt{8\pi c^2 / \int_V \left[ \frac{d(\omega\varepsilon)}{d\omega} + \varepsilon \right] \mathbf{g}_\omega . \mathbf{g}_\omega^* d^3\mathbf{r}} \quad . \tag{3}$$

This gives $\langle \mathcal{H}_\omega \rangle = \omega^2 Q_\omega^2 / 2$, which has the form of the average energy of a harmonic oscillator whose coordinate is $Q_\omega$ [15]. Using the rigorous result (valid for the empty cavity as well as for homogeneous or inhomogeneous dielectric media) that the Lagrangian involving fields reduces to that of a set of independent classical oscillators [13 - 19], it follows that $Q_\omega$ and the canonical oscillator coordinate must be one and the same. Considering all the modes, the classical Hamiltonian is therefore $\mathcal{H} = \sum_s \left(P_s^\dagger P_s + \omega_s^2 Q_s^\dagger Q_s\right)/2$, where $s = 1, 2, ..\infty$ is the mode index ($\omega_1 < \omega_2 < ..$) and $P_s$ is the momentum conjugate to $Q_s$. The modal solutions satisfy the orthogonality condition $\int_V \left[\omega_s^2 \varepsilon(\mathbf{r}, \omega_s) - \omega_t^2 \varepsilon(\mathbf{r}, \omega_t)\right] \mathbf{g}_{\omega_s}(\mathbf{r}).\, \mathbf{g}_{\omega_t}^*(\mathbf{r}) d^3\mathbf{r} = 0$ ($s \neq p$) [14,18] and can be normalized to give $C_\omega \equiv 2\pi^{1/2} c$ for all eigenfrequencies (since $v << V$, the required normalization condition is approximately $\int_V \mathbf{g}_\omega . \mathbf{g}_\omega^* d^3\mathbf{r} = 1$). Thus, the classical field is

$$\mathbf{A}_Q(\mathbf{r}, t) = 2\pi^{1/2} c \sum_s Q_s \mathbf{g}_{\omega_s}(\mathbf{r}) e^{-i\omega_s} \quad . \tag{4}$$

The analogous expression for the empty cavity is

$$\mathbf{A}_U(\mathbf{r}, t) = 2\pi^{1/2} c \sum_s U_s \mathbf{f}_{\Omega_s}(\mathbf{r}) e^{-i\Omega_s} \tag{5}$$

where $\mathbf{f}_{\Omega_s}$ $U_s$ and $\Omega_s$ denote, respectively, the vector field, coordinate and eigenfrequency of a particular mode, with $\int_V \mathbf{f}_{\Omega_s}(\mathbf{r}).\, \mathbf{f}_{\Omega_p}^*(\mathbf{r}) d^3\mathbf{r} = \delta_{sp}$. We recall that the set $\{\mathbf{f}_\Omega\}$ is complete, that is, an arbitrary field inside the cavity can be expressed as a sum over all the modes.

To quantize the model, we replace $Q_\omega$ and $U_\Omega$ with the corresponding quantum operators in the Schrödinger picture, or with $i\sqrt{\hbar / 2\omega}\left(a_\omega - a_\omega^\dagger\right)$ and $i\sqrt{\hbar / 2\Omega}\left(a_\Omega - a_\Omega^\dagger\right)$ where $a_\omega^\dagger$ ($a_\Omega^\dagger$) and $a_\omega$ ($a_\Omega$) are the dressed (bare) photon creation and annihilation operators. The associated canonically conjugated operators are given by the well-known expressions $-i\hbar\partial / \partial Q_\omega$ and $-i\hbar\partial / \partial U_\Omega$. Using (4) and (5), and assuming that the set $\{\mathbf{g}_\omega\}$ is also complete [20], we obtain the following, linear relationships involving the two coordinate sets

$$U_s = \sum_p D_{sp} Q_p \qquad Q_p = \sum_s D^{-1}_{ps} U_s \tag{6}$$

where

$$D_{sp} = \int_V \mathbf{f}^*_s(\mathbf{r}).\, \mathbf{g}_p(\mathbf{r})\, d^3\mathbf{r} \quad . \tag{7}$$

Note that, for dispersionless media, $\int_V \varepsilon(\mathbf{r}) \mathbf{g}_{\omega_s}(\mathbf{r}).\, \mathbf{g}^*_{\omega_p}(\mathbf{r}) d^3\mathbf{r} = \delta_{sp}$ [10] and, thus, $D^{-1}_{ps} = \int_V \varepsilon(\mathbf{r}) \mathbf{f}_s(\mathbf{r}).\, \mathbf{g}^*_p(\mathbf{r})\, d^3\mathbf{r}$. It is apparent that the completeness of the set $\{\mathbf{g}_\omega\}$ is tantamount to the existence of the inverse matrix $D^{-1}_{ps}$.

We now have all the ingredients to calculate the overlap between the two ground states: $|0_\Omega\rangle$ (empty cavity) and $|0_\omega\rangle$ (with inclusions). To that end, we use the familiar ground-state wave-function of a harmonic oscillator to calculate the partial overlap

$$S(N) = \frac{\int_{-\infty}^{+\infty} e^{-\frac{1}{2}\sum_{i=1}^{N}(\Omega_i U_i^2 + \omega_i Q_i^2)} dQ_1 \ldots dQ_N}{\left\{ \int_{-\infty}^{+\infty} e^{-\sum_{i=1}^{N}\Omega_i U_i^2} dQ_1 \ldots dQ_N \right\}^{1/2} \left\{ \int_{-\infty}^{+\infty} e^{-\sum_{i=1}^{N}\omega_i Q_i^2} dQ_1 \ldots dQ_N \right\}^{1/2}} \quad , \tag{8}$$

defined as the overlap between states corresponding to the first $N$ cavity modes. Clearly, $\langle 0_\Omega | 0_\omega \rangle = S(N \to \infty)$. Introducing the symmetric matrix $C_{sp} = \sum_{j=1,N} \Omega_j D_{js} D_{jp}$, and using the Jacobian (= $\det_N |D_{sp}|$) for the change of variables, we finally obtain

$$S(N) = 2^{(N+1)} \left( \prod_{i=1}^{N} \Omega_i^{1/4} \omega_i^{1/4} \right) \sqrt{\frac{\det_N |D_{sp}|}{\det_N |\omega_s \delta_{sp} + C_{sp}|}} \tag{9}$$

where $\det_N |D_{sp}|$ comprises overlaps associated with the first $N$ modes, that is, $s, p = 1, .., N$ in Eq. (7). Some reflection shows that $\det_N |D_{sp}|$ can be interpreted in terms of the many-body overlap between two Slater determinants representing the unperturbed and locally-perturbed ground states of a system of $N$ free electrons, which is known to be of order $N^{-\eta}$ $(\eta > 0)$ in the thermodynamic limit [2]. Central to our contention that perturbations due to small polarizable particles can lead to orthogonality catastrophes, this mode-to-wavefunction mapping of overlaps defines the close relationship that exists between the electron and photon problems, notwithstanding obvious differences in regard to boundary conditions, the vector vs. scalar and the bosonic vs. fermionic nature of the states [21].

In the following, we apply the general theory to a cavity delimited by a perfectly-conducting spherical shell of radius $R$, which contains a concentric sphere of radius $a << R$, whose permittivity is $\varepsilon_S(\omega)$. Solutions divide into transverse-electric (TE) and transverse-magnetic (TM) modes and can be found exactly [ 22 ]. In particular, $\mathbf{g}_{lm} = \mathbf{X}_{lm}(\theta,\varphi) g_l(r)$ with $\mathbf{X}_{lm} = -i(\mathbf{r}\times\nabla)Y_{lm} / \sqrt{l(l+1)}$ for TE modes ($Y_{lm}$ are spherical harmonics). Using the requirement that the electric field vanish at $r = R$, and the continuity of the electric field and the tangential component of the magnetic field at $r = a$, we obtain the unnormalized solutions

$$g_l(r) = \begin{cases} j_l(n_S kr) & r < a \\ j_l(n_S\beta)\dfrac{j_l(kr)y_l(kR) - y_l(kr)j_l(kR)}{j_l(\beta)y_l(kR) - y_l(\beta)j_l(kR)} & r \geq a \end{cases} \tag{10}$$

and the equation giving the resonant wavevectors

$$\frac{j_l(n_S\beta)}{[n_S\beta j_l(n_S\beta)]'} = \frac{j_l(\beta)y_l(kR) - y_l(\beta)j_l(kR)}{[\beta j_l(\beta)]' y_l(kR) - [\beta y_l(\beta)]' j_l(kR)} \quad . \tag{11}$$

Here, $k = \omega / c$ ($c$ is the speed of light in vacuum), $\beta = ka$, and $n_S = \sqrt{\varepsilon_S}$; $j_l$ ($y_l$) is the spherical Bessel function of the first (second) kind of order *l*. The corresponding expressions for TM modes are easily derived [22]. For the empty cavity, the unnormalized TE solutions are simply $f_l(r) = j_l(qr)$, where $q = \Omega / c$, while $j_l(qR) = 0$ gives the eigenfrequencies.

Because of the symmetry of the problem, the single-function overlaps entering $D_{sp}$, Eq. (7), vanish unless the two states share the same *l* and *m*. Hence, $D_{sp}$ divides into separate blocks identified by specific values of these quantum numbers. Within a block, overlaps can be straightforwardly obtained using the asymptotic form of the spherical Bessel functions $f_l \approx \sin(qr - l\pi/2)/qr$ and $g_l(r) \approx \sin[k(r-R)]/kr$, valid for $qr >> l$. In particular, for odd values of *l*,

$$\langle k | q \rangle = \frac{2k(\cos kR - \cos qR)}{(k^2 - q^2)R\sqrt{\left(1 - \frac{\sin 2kR}{2kR}\right)\left(1 - \frac{\sin 2qR}{2qR}\right)}} . \tag{12}$$

Central-cell corrections accounting for the differences between the exact and the asymptotic- form overlaps are not important in the limit $R \to \infty$. Finally, we recall that the eigenvalues for the two problems are related through $k_t = q_t + \delta_l / R$, where $\delta_l(q_t)$ is the scattering phase shift, which can be gained without difficulty from the Mie coefficients buried in Eq. (10) [23].

The above discussion has not yet revealed the anticipated orthogonality catastrophe, except for a brief comment on the relationship between $\det_N |D_{sp}|$ and overlaps of electron Slater determinants. To do so, we examine the problem of a sphere made of a metal that obeys Drude's formula $\varepsilon_S(\omega) = 1 - \omega_P^2 / \omega^2 = 1 - k_P^2 / k^2$, where $\omega_P$ $(= ck_P)$ is the plasma frequency. For simplicity, we consider from now on only $l = 1$ TE states for which the resonant wavevectors of the empty cavity obey $\tan(qR) = qR$ [24]. The results in Fig. 1 reveal the orthogonality catastrophe. The contour plot, Fig. 1(a), shows calculated values of $\det_N |D_{sp}|$ at $k_{\rm P} = 5/a$ as a function of *N* and *R*, while Fig. 1(b) both reproduces the contour data and shows $S^2(N,R)$ along the line $Na/R = 1.88$ where the determinant is smallest for fixed *N* or *R*. The calculations were performed using Eq. (12) for the single-mode overlaps and the exact resonant wavevectors of the empty cavity. The $l = 1$ phase shift, gained from Eq. (10) and well-known expressions from scattering theory [23], was used to obtain the corresponding wavevectors for the cavity containing the Drude sphere; see below. The linear fit to the determinant data in Fig. 1(b) translates into $\det_N |D_{sp}| \propto N^{-0.39}$. $S^2$ decreases with *N* with roughly the same exponent.

The calculated $l = 1$ TE phase shift, $\delta_1$, is shown in Fig. 2. The main peak occurs slightly above $k_{\rm P}$ whereas the other features are due to Fabry-Pérot- like resonances at integer multiples of $\pi/a$. We find that $\delta_1 \propto 1/k$ for $k \to \infty$ [24] while, as expected, $\delta_1 \propto k^2$ for $k \to 0$. Note that a Drude metal behaves as a perfect mirror for $k < k_{\rm P}$ where the refractive index is purely imaginary. The value of $Na/R$ in Fig. 1(b) corresponds to the wavevector $k \approx 1.18 k_{\rm P}$ at which the phase shift is a maximum. It should be noted that the peak height in Fig. 2 increases with increasing $k_{\rm P}$ and that it can attain values larger than $\pi$.

The results of Fig. 1 as well as calculations for many other values of the parameters indicate that the determinant of $D_{sp}$ controls the behavior of *S* at large *N* and that, at a given value of $Na/R$, $\det_N |D_{sp}| \sim N^{-\eta}\ (\eta > 0)$. Moreover, the dependence of the exponent $\eta$ on $Na/R$ closely follows that of the phase shift on the wavevector $k = N\pi / R$. Note, in particular, the strong asymmetry with respect to the line $Na / R = 1.88$ in Fig. 1(a), which faithfully reproduces the asymmetry of the phase shift with respect to the peak at $k \approx 1.18 k_{\mathrm{P}}$; see Fig. 2. Since $\delta_1(k) \neq 0$, except at $k = 0$ and $k = \infty$, this means that, other than for $N \equiv \text{constant}$ and $R \equiv \text{constant}$, the states become orthogonal in the infinite limit for arbitrary values of $Na / R$.

The behavior of the electromagnetic field vis-à-vis the insertion of a polarizable particle, especially the power-law decrease of the overlap with *N* and the dependence of the exponent on the phase shift at $Na / R$, strongly resembles that of a system of electrons perturbed by a local potential [2,4]. More precisely, the photon problem for the first *N* modes of a cavity of radius *R* relates to that of a system of *N* free electrons with Fermi wavevector $k_{\mathrm{F}} = N\pi / R$. This mapping of overlaps, alluded to earlier, is a key result which allows us to make predictions for the electromagnetic field based on what is already known from electron studies. In particular, the fact that the exponent $\eta$ depends only on the scattering phase shift strongly suggests that the catastrophe is a general phenomenon, not limited to Drude-type spherical inclusions. Also, since an exceedingly small phase shift leads to orthogonality, the ground states with and without inclusions cannot be adiabatically connected in the infinite limit because their overlap changes abruptly from one to zero, regardless of how close the inclusion's permittivity is to the vacuum's value. We further recall that the Fermi ground state of the perturbed system is not only orthogonal to the unperturbed ground state, but to all states containing a finite number of electron-hole excitations [2]. Since the total energy change is finite when a local potential is added, it follows that its insertion must be accompanied by a divergence in the number of excitations as their energy approaches zero. This is the infrared divergence mentioned in the introduction which, by analogy, should reveal itself in the creation of a macroscopic number of low-frequency photons when a polarizable particle is inserted in or removed from a cavity. Another important aspect of the catastrophe is that it also applies to the overlap between states corresponding to different positions of the local potential [25]. In photon terms, this means that the displacement of a polarizable particle inside a large cavity must also result in the creation of a diverging number of low frequency photons. This prediction, which bears on the dynamical Casimir effect, that is, the generation of photons from vacuum due to the motion of uncharged boundaries [26,27,28], can be tested experimentally.

Finally, we comment briefly on the possible relevance of these results to the quantum measurement problem. In [29], we argue that there are only two types of measuring devices involving (*i*) phase transformations (*e. g.*, the bubble chamber) or (*ii*) macroscopic transfers of charge (*e. g.*, the Geiger counter). It is apparent that, by locally changing the permittivity or the boundary conditions on the electric field, a single measurement with either class of devices perturbs the electromagnetic modes (of the universe!) as much as the insertion of a polarizable particle perturbs a cavity. Hence, the effect of a measurement on the photon Hilbert space is that of a transformation leading to a unitarily inequivalent representation. One could then argue, as done in [29], that coherent superpositions of the Schrödinger’s cat type cannot be allowed, since they violate the uniqueness of the Hamiltonian. Within this context, and given that the infinite limit extends beyond the range of frequencies where electrodynamics of continuous media applies, it would be of interest to widen our studies to x-ray and gamma-ray frequencies.

To summarize, we presented arguments and numerical calculations uncovering catastrophic effects caused by the insertion of a small polarizable object in a large electromagnetic cavity, thereby revealing the existence of a mapping from the photon problem to that of a many-body system of electrons perturbed by a local potential. Using this relationship, we made the prediction that the insertion, removal or displacement of a polarizable particle must be accompanied by the production of photons with a diverging distribution at low frequencies.

Work supported by the MRSEC Program of the NSF under Grant No. DMR-1120923.

## REFERENCES

[1] See, e.g., N. N. Bogolubov, A. A. Logunov, A. I. Oksak and I. T. Todorov, *General Principles of Quantum Field Theory* (Kluwer, Dordrecht, 1990), Ch. 6.

[2] P. W. Anderson, "Infrared catastrophe in Fermi gases with local scattering potentials," Phys. Rev. Lett. **18**, 1049 (1967).

[3] G. D. Mahan, *Many-Particle Physics*, 3rd edition, (Kluwer Academic/Plenum Publishers, New York, 2000), Ch. 9.

[4] K. Ohtaka and Y. Tanabe, "Theory of the soft x-ray problem in simple metals: historical survey and recent developments," Rev. Mod. Phys. **62**, 929 (1990).

[5] k. Yosida and A. Yoshimori, in *Magnetism*, ed. by H. Suhl (Academic Press, New York, 1973), Vol. **V**, p. 25.

[6] J. J. Hopfield, "Infrared divergences, x-ray edges, and all that," Comments Solid State, Phys. **2**, 40 (1969).

[7] C. B. Duze and G. D. Mahan, "Phonon-Broadened Impurity Spectra. I. Density of States," Phys. Rev **9**, A1965 (1965).

[8] A. A. Louis and J. P. Sethna, "Atomic tunneling from a scanning-tunneling or atomic-force microscope tip: dissipative quantum effects from phonons," Phys. Rev. Lett. **74**, 1363 (1995).

[9] L. D. Landau and E. M. Lifshitz, *Electrodynamics of Continuous Media* (Pergamon, Oxford, 1960).

[10] R. J. Glauber and M. Lewenstein, “Quantum optics of dielectric media,” Phys. Rev. A **43**, 467 (1991).

[11] L. Knöll and D. G. Welsch, “QED of resonators in quantum optics,” Prog. Quantum Electron. **16**, 135 (1992).

[12] S. Scheel and D. G. Welsch, “Interaction of the quantized electromagnetic field with atoms in the presence of dispersing and absorbing dielectric bodies,” Opt. Spectrosc. **91**, 508 (2001).

[13] P. D. Drummond, “Electromagnetic quantization in dispersive inhomogeneous nonlinear dielectrics,” Phys. Rev. A **42**, 6845 (1990).

[14] D. J. Santos and R. Loudon, “Electromagnetic-field quantization in inhomogeneous and dispersive one-dimensional systems,” Phys. Rev. A **52**, 1538 (1995).

[15] P. W. Milonni, “Field quantization and radiative processes in dispersive dielectric media,” J. Mod. Opt. **42**, 1991 (1995).

[16] B. Huttner, J. J. Baumberg and S. M. Barnett, “Canonical quantization of light in a linear dielectric,” Europhys. Lett. **16**, 177-182 (1991).

[17] R. Matloob, R. Loudon, S. M. Barnett and J. Jeffers, “Electromagnetic field quantization in absorbing dielectrics,” Phys. Rev. A **52**, 4823 (1995).

[18] Z. Lenac, “Quantum optics of dispersive dielectric media,” Phys. Rev. A **68**, 063815 (2003).

[19] A. C. Judge, M. J. Steel, J. E. Sipe and C. M. de Sterke, “Canonical quantization of macroscopic electrodynamics in a linear, inhomogeneous magnetoelectric medium,” Phys. Rev. A **87**, 033824 (2013).

[20] Completeness of a modal set follows from the equal-time commutation relation $\left[\mathbf{A}(\mathbf{r},t),\boldsymbol{\Pi}(\mathbf{r}',t)\right]=i\hbar c^{-1}\delta(\mathbf{r}-\mathbf{r}')$ involving the conjugated fields $\mathbf{A}$ and $\boldsymbol{\Pi}=-c^{-1}\varepsilon\mathbf{E}$ [12,13,14,16,17,18]. As discussed in [12,14,17], a rigorous proof of the validity of the closure relation for mode functions requires the consideration of the complex nature of the permittivity, in accordance to the Kramers-Kronig relations.

[21] A perfect conductor provides a simple example of orthogonality catastrophe because their modal functions vanish inside the inclusions and, thus, $\lim_{N\to\infty}\det_N\left|D_{sp}\right|=0$ for all *R;* numerical calculations give $S\approx(\pi/2)^{-(Na/R)^2}$. However, note that the inherent assumption that the permittivity diverges in a certain range of frequencies violates causality requirements.

[22] J. D. Jackson, *Classical Electrodynamics*, 3rd Edition (Wiley, New York, 1999), Ch. 9.

[23] See, e.g., H. C. van de Hulst, *Light scattering by small particles* (Dover, New York, 1981).

[24] An important difference between TE and TM states is that the latter alone allow plasmon resonances, and these lead to phase shifts which approach π at large wavevectors.

[25] K. Schönhammer, "Orthogonality exponent and the friction coefficient of an electron gas," Phys. Rev. **43**, 11323 (1991).

[26] J. Schwinger, "Casimir light: A glimpse," Proc. Nat. Acad. Sci. USA **90**, 958-959 (1993).

[27] V. V. Dodonov, "Dynamical Casimir effect: some theoretical aspects," J. Phys.: Conf. Ser. **161**, 012027 (2009).

[28] A. Belyanin, V. V. Kocharovsky, F. Capasso, E. Fry, M. S. Zubairy and M. O. Scully, "Quantum electrodynamics of accelerated atoms in free space and in cavities," Phys. Rev. A **74**, 023807 (2006).

[29] R. Merlin, “A heuristic approach to the quantum measurement problem: How to distinguish particle detectors from ordinary objects,” Int. J. Mod. Phys. B **29**, 1530011 (2015).

FIGURE CAPTIONS

FIG. 1 (color online). Drude sphere of radius *a* inside a cavity of radius *R*. Data for $l = 1$ TE modes at $k_{\mathrm{P}} a = 5$. (a) Contour plot of $\det_N |D_{sp}|$ as a function of *R/a* and the number of modes *N*. (b) *N*-dependence of $\det_N |D_{sp}|$ and square of the partial ground-state overlap for $Na / R = 1.88$.

FIG. 2. Wavevector dependence of the $l = 1$ phase shift for TE modes, $\delta_1$ (units of $\pi$) for a Drude sphere; $k_{\mathrm{P}} = 5 / a$ is the plasma wavevector.

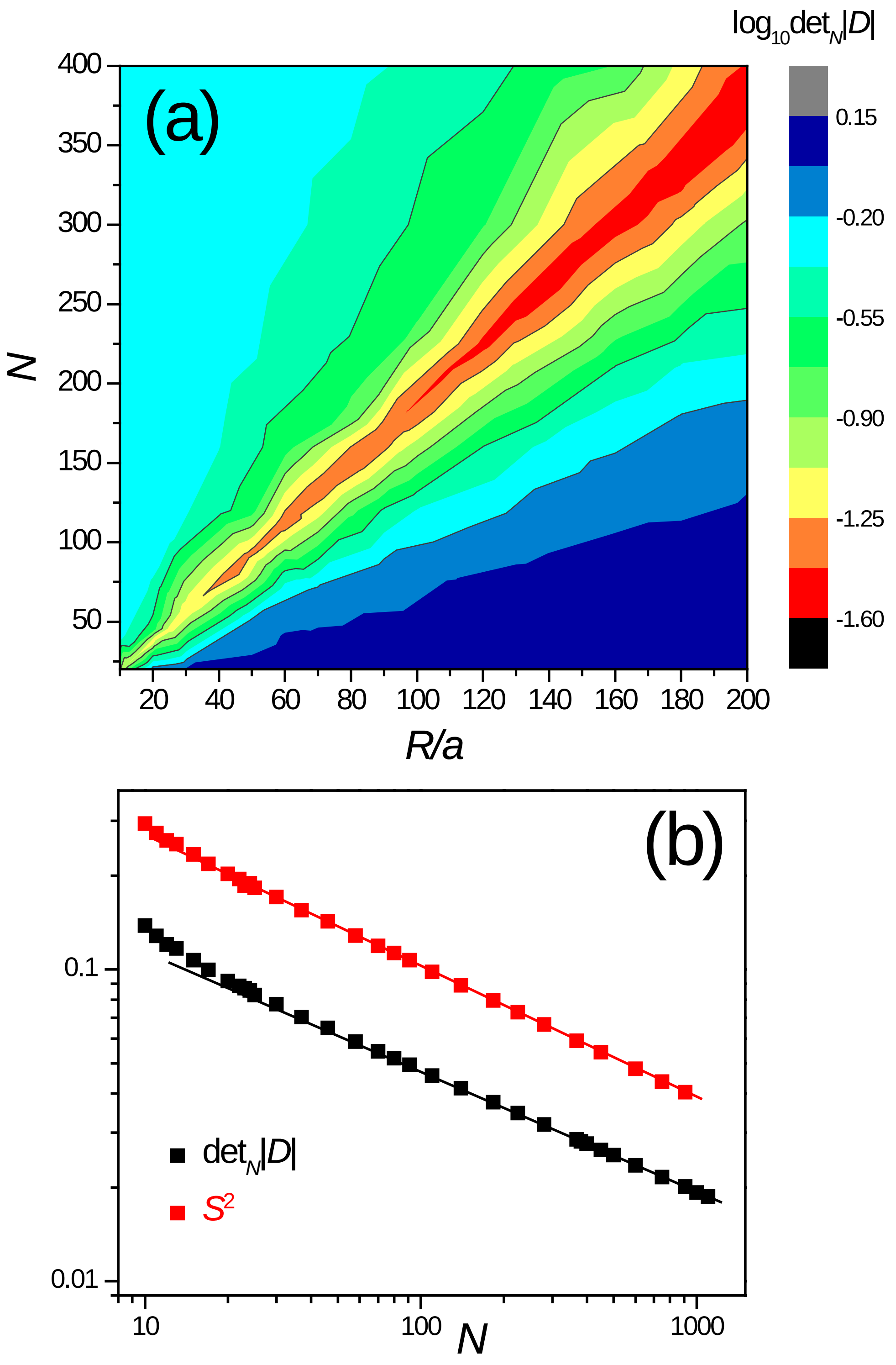

(a)
log10detN|D|
0.15
-0.20
-0.55
-0.90
-1.25
-1.60
N
400
350
300
250
200
150
100
50
20 40 60 80 100 120 140 160 180 200
R/a
(b)
0.1
0.01
10 100 1000
N
detN|D|
S2

FIGURE 1

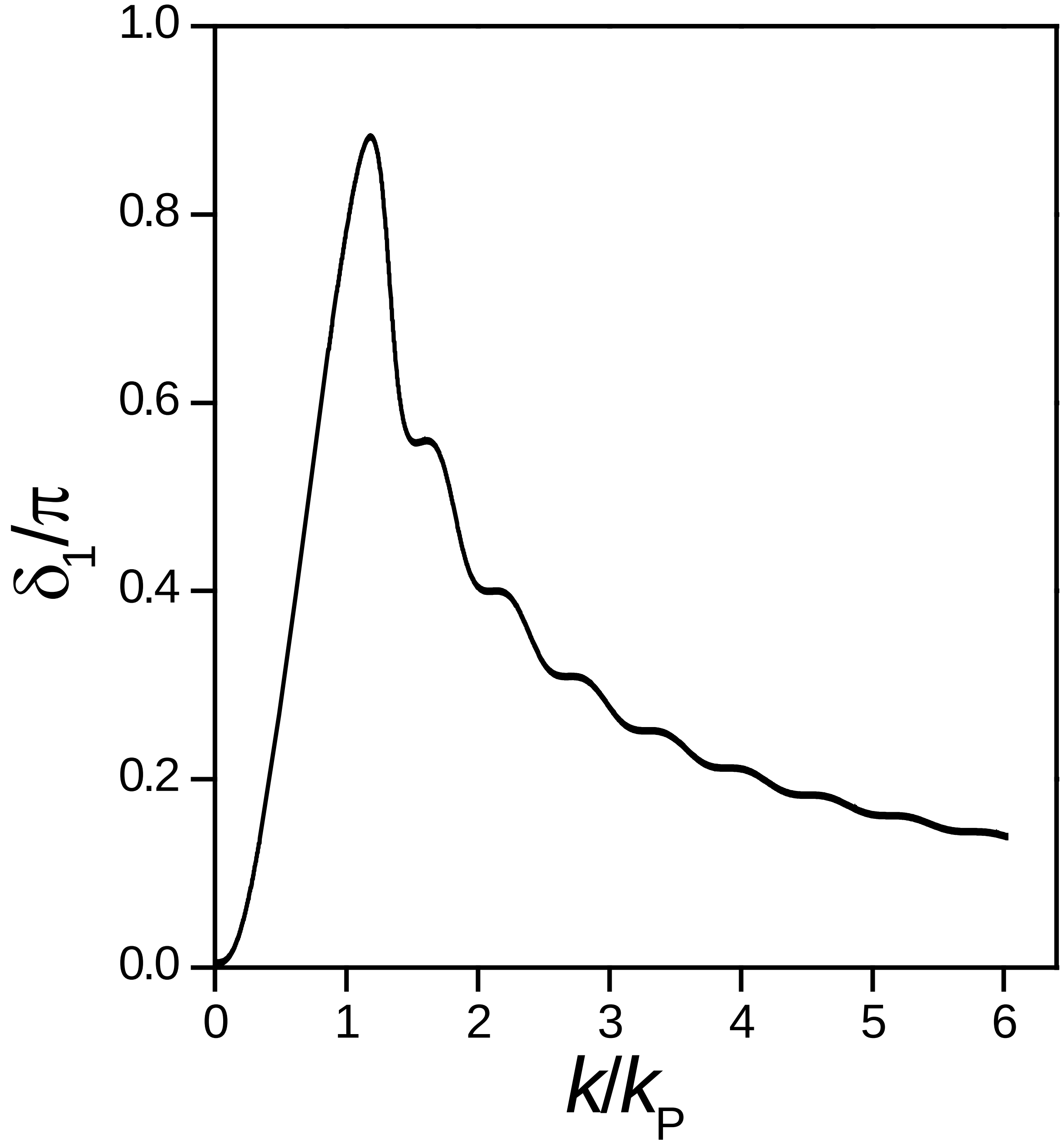

FIGURE 2